\documentclass[epj,nopacs,onecollarge]{svjour}

\usepackage{graphicx}
\usepackage{subfigure}
\usepackage{color}
\usepackage{cases}
\usepackage{float}
\usepackage{hyperref}
\usepackage{breakurl}
\usepackage{epstopdf}
\usepackage{amssymb}

\begin{document}

\title{Comparing numerical and analytical approaches to strongly interacting two-component mixtures in one dimensional traps}

\titlerunning{Comparing numerical and analytical approaches to strongly interacting systems}

\author{Filipe~F.~Bellotti \and Amin~S.~Dehkharghani \and Nikolaj~T.~Zinner}

\institute{Department of Physics and Astronomy, Aarhus University, DK-8000 Aarhus C, Denmark \\ \email{bellotti@phys.au.dk} }


\date{Received: date / Revised version: date}

\maketitle

\begin{abstract}
We investigate one-dimensional harmonically trapped two-component systems for repulsive interaction strengths ranging from the non-interacting to the strongly interacting regime for Fermi-Fermi mixtures. 
A new and powerful mapping between the interaction strength parameters from a continuous Hamiltonian and a discrete lattice Hamiltonian is derived. As an example, we show that this mapping does not depend neither on the state of the system nor on the number of particles. 
Energies, density profiles and correlation functions are obtained both numerically (DMRG and Exact diagonalization) and analytically. 
Since DMRG results do not converge as the interaction strength is increased, analytical solutions are used as a benchmark to identify the point where these calculations become unstable. We use the proposed mapping to set a quantitative limit on the interaction parameter of a discrete lattice Hamiltonian above which DMRG gives unrealistic results. 
\end{abstract}

\section{Introduction}
One-dimensional (1D) systems are among the most widely studied problems in physics, especially due to their invaluable pedagogical properties and their more friendly manipulation of mathematical expressions both analytically and numerically, which often guide us through the understanding of interesting physical systems \cite{TonksPR1936,GirardeauJoMP1960,blochRMP2008,lewensteinAiP2007,esslingerARoCMP2010,Giamarchi2003,sowinskiPRA2013,VolosnievPRA2015,HuNJoP2016,YangPRA2016,PecakNJoP2016,YangPRA2015,CampbellPRA2014,Garcia-MarchNJoP2014,Garcia-MarchPRA2015,Garcis-MarchJoPBAMaOP2016,ZoellnerPRA2008,TempfliNJoP2009,BrouzosPRL2012,BrouzosPRA2013,DeuretzbacherPRL2008,SowinskiEEL2015}. 
Furthermore, 1D structures such as nanotubes and nanowires, among others, may be highly relevant in technological applications \cite{Altomare2013}. Beyond that, such low-dimensional systems can be realized in experiments with cold atomic gases \cite{ParedesN2004,KinoshitaN2006,MurmannPRL2015a,HallerS2009,PaganoNP2014,KinoshitaS2004} even in the limit of few-particles \cite{SerwaneS2011,ZuernPRL2013,MurmannPRL2015}. Combined with the tunable interaction between atoms \cite{GreinerN2002,WenzS2013,zinnerJoPGNaPP2013,zurnPRL2012,chinRMP2010,OlshaniiPRL1998} they form the perfect test-bed for studying the quantum mechanics of few- and many-body systems \cite{ZinnerEWoC2016,HofmannPRA2016}.

A key feature of 1D systems is that analytical solutions of many-body physics are known, such as for instance those obtained via the Bethe ansatz. There are, unfortunately, not many solutions available when particles are confined in 
external traps as is the case of recent experiments
\cite{SerwaneS2011}. More generally, 1D systems of many interacting particles have been
studied numerically and a powerful method in use is the density matrix renormalization group (DMRG) \cite{WhitePRL1992}. DMRG codes based on matrix product states (MPS) (e.g. \cite{Wall2009,WallNJoP2012,itensor}) are accurate, fast and have been successfully used to study real-time evolutions either at zero \cite{VidalPRL2003,BanulsPRL2009} or finite temperature \cite{FeiguinPRB2005}, to efficiently implement periodic boundary conditions \cite{PippanPRB2010}, in numerical renormalization group applications \cite{WeichselbaumPRB2009}, in development of infinite-system algorithms \cite{VidalPRL2007} and in several other contexts.

The success of DMRG \cite{DeChiaraJoCaTN2008,SchollwoeckRMP2005,SchollwoeckJoMaMM2007} has led the method to be pushed to limiting cases such as the study of continuous systems without a lattice parameter \cite{VerstraetePRL2010}. This limit is also accessed from a discrete prescription, in which case it is only valid when the occupation level $N$ is much less than the number of lattice sites $L$, namely $N/L<<1$. The way to respect this relation is to take the limit of large $L$ for a fixed number of particles ($N>1$), which calls for extra care since the method is not expected to work in the extreme limit $L\to\infty$ \cite{SchollwoeckJoMaMM2007}. Another example is the application of DMRG methods in the limit of strong interactions \cite{FangPRA2011}.

Here we are interested in how well DMRG performs in dealing with continuous systems in the strongly interacting limit. Interest in this setup goes beyond its experimental realization \cite{HallerS2009,PaganoNP2014,KinoshitaS2004,SerwaneS2011,GreinerN2002}, and arises also from the fact that an exact solution for such systems is available \cite{VolosnievNC2014,DeuretzbacherPRA2014}. This important result serves as benchmark for our investigation since it provides a precise reference for numerical calculation. 
We propose a practical and efficient way to connect parameters from continuous and discrete Hamiltonians, which may be a relevant tool for studying continuous systems through lattice Hamiltonians. This new mapping allows us to set a quantitative limit on the interaction parameter of a discrete lattice Hamiltonian above which DMRG gives unrealistic results. 

The structure of the paper is as follows. Sect. \ref{sec.sm} present our systems and the methods and in
Sect. \ref{sec.dcm} we present the new mapping between parameters from continuous and discrete Hamiltonians.
The behavior of the DMRG method is investigated in Sect. \ref{sec.sisdmrg} for two-component fermions and results for energies, density profiles and pair correlation functions are discussed. 
Concluding remarks are given in Sect. \ref{sec.conc}.

\section{System and methods} \label{sec.sm}
We consider a two-component system composed of $N=N_a+N_b$ particles whose dynamics is restricted to one spatial dimension. Each component $a$ and $b$ can be either a fermion or a boson in a specific internal state. The $N$ particles have mass $m$ and components $a$ and $b$ are distinguished from each other only by their internal state. All particles are confined to the same one-dimensional harmonic trap $V(x)$ and the short-range interaction between pairs is taken to be only repulsive. In the following subsections we present three ways to describe and investigate this system.

\subsection{Continuous description}
This is the description that straightforwardly connects theoretical and experimental results \cite{SerwaneS2011,andersenSR2016}. The one-dimensional harmonic trap acting on each particle reads $V(x)=m\omega^2x^2/2$. Pairwise interaction is modeled through a Dirac-$\delta$ function as $U_{ij}(x_j-x_i)=g_{ij} \delta(x_j-x_i)$ with $g_{ij}\geq0$. The interaction strength is $g_{ij}=g$ if $i,j$ are either different species or identical bosons allowed to interact and $g_{ij}=0$ when $i$ and $j$ are fermions from the same species. The continuous Hamiltonian describing the system is given by
\begin{equation} 
H_c=\sum_{i=1}^{N} \left( -\frac{\hbar^2 }{2 m} \frac{\partial^2}{\partial x_i^2} + \frac{m \omega^2}{2} x_i^2 \right) + \sum_{j<i}^{N} g_{ij} \delta(x_j-x_i) ,
\label{hamil}
\end{equation} 
where the second term on the right-hand-side implies that any wave function $\psi(x_1,...,x_N)$ obeys the boundary condition
\begin{equation}
\left.\left(\frac{\partial \psi}{\partial x_i}-\frac{\partial \psi}{\partial x_j} \right)\right|_{x_i-x_j\to 0^+}-\left.\left(\frac{\partial \psi}{\partial x_i}-\frac{\partial \psi}{\partial x_j} \right)\right|_{x_i-x_j\to 0^-}  =  
\frac{2m}{\hbar^2} g_{ij} \psi\left( x_i=x_j \right) . 
\label{bc}
\end{equation}

The Hamiltonian Eq. \ref{hamil} is used to solve the system with the exact diagonalization technique \cite{DehkharghaniSR2015}, variational method \cite{andersenSR2016} and together with the boundary condition Eq. \ref{bc} to obtain the exact analytical wave function in the strongly interacting limit \cite{VolosnievNC2014}, achieved when $g$ is much larger than all other scales in the system. The exact diagonalization is done by considering an effective two-body interaction, which speeds up the convergence even for large values of $g$. More details about the construction of the effective potential in the truncated two-body space for the exact diagonalization is given in \cite{DehkharghaniSR2015}.

\subsubsection{Analytical solutions in the strongly interacting limit}
Only the relevant quantities for this work are presented here and the interested reader can find more details about their derivation in \cite{VolosnievNC2014,VolosnievFS2014,decampNJoP2016}. 
It is assumed that all particles interact with the same zero-range interaction of strength $g$, irrespective of the statistics they obey and whatever the internal state is. Note that for identical fermions this just means no interaction at all due to the Pauli principle, which requires antisymmetry under exchange of two such identical fermions. In the strongly interacting limit $1/g\to0$, the eigenstate wave function of a two-component system composed of $N=N_a+N_b$ particles is written as

\begin{numcases}
{\psi(x_1,...,x_N)= \label{psiana}}
a_1 \Psi_A & for $x_1<...<x_{N_a}<x_{N_a+1}<...<x_N$ $(a...ab...b)$ \nonumber \\
a_2 \Psi_A & for $x_1<...<x_{N_a+1}<x_{N_a}<...<x_N$ $(a...ba...b)$ \nonumber \\
\vdots & $\vdots$ \\
a_M \Psi_A & for $x_N<...<x_{N_a+1}<x_{N_a}<...<x_1$ $(b...ba...a)$ \nonumber 
\end{numcases}

where $x_n$ is the coordinate of the $n^{th}$ particle and $M=N!/(N_a! N_b!)$ is the number of independent distinguishable spatial configurations or in other words $M$ is the number of degenerated states at $1/g\to0$. The wave function $\Psi_A$ is constructed from the antisymmetric product of the first $N$ eigenstates of the single-particle continuous Hamiltonian (the first term in Eq. \ref{hamil}) and it has energy $E_A$ which is the sum of the occupied single-particle energies. 

Up to linear order, the energy of the system in this limit can be written as $E=E_A-K/g$, where $K=K(a_1,...,a_M)$ depend on the $M$ coefficients of the wave function Eq. \ref{psiana} and it is independent of $g$, namely
\begin{equation}
K(a_1,...,a_M)=\sum_{k,p=1}^M (a_k-a_p)^2 I_{k,p} \ ,
\label{slopes}
\end{equation} 
with $\langle \psi | \psi \rangle=1$ and coefficients $I_{i,j}$ given by
\begin{equation}
I_{i,j}=\int_{x_1<...<x_N} dx_1...dx_N \delta\left(x_i-x_j\right) \left| \frac{\partial \Psi_A}{\partial x_i} \right|^2 \ .
\end{equation}
The slope $K$ for each state and the respective $a_k$ coefficients are found by diagonalizing the system of linear equations obtained from the variation of equation Eq. \ref{slopes} with respect to $a_k$, i.e. $\partial K/\partial x_k=0$, for $k \in \{1,...,M\}$.

To get the one-body density of the component with index $\mu$ we must calculate 
\begin{equation}
n_{\mu}(x)=N_\mu \int dx_1...dx_N \delta\left(x-x_\mu\right) \left| \psi(x_1,...,x_N) \right|^2 \ ,
\end{equation}
where $\mu=a,b$ and $N_\mu=N_a,N_b$. 
Similarly, the pair correlation function relates how the different species are spatially organized with respect to each other in the trap. It has the explicit formula
\begin{equation}
\Sigma\left(x,y\right)= \int dx_1...dx_N \delta\left(x-x_1\right) \delta\left(y-x_N \right) \left| \psi(x_1,...,x_N) \right|^2.
\label{paircoreq}
\end{equation}
Since we consider two-component systems here we always have at least two particles of different kind. This means that the construction of the wave function according to equation Eq. \ref{psiana} will always ensure that the coordinates $x_1$ and $x_N$ belong to different species. The pair correlation function is therefore an inter-species pair correlation measure.

\subsection{Discrete description}
The discrete modeling of physical systems is often used in the study of many condensed-matter systems including for instance spin chains \cite{FuehringerAdP2008,MorigiPRL2015}. Generally, we observe that this method resembles a continuous system when the occupation density is low, namely $N/L\ll1$, with $N$ being the number of particles and $L$ the number of discrete lattice sites. 
The lattice discretization of Eq. \ref{hamil} will hereafter be referred as the discrete Hamiltonian, $H_d$, and reads

\begin{eqnarray} 
H_d & =  -t \sum_{j=1}^{N-1} \left( a_{j}^{\dagger} a_{j+1} + a_{j+1}^{\dagger} a_{j} \right) 
-t \sum_{j=1}^{N-1} \left( b_{j}^{\dagger} b_{j+1} + b_{j+1}^{\dagger} b_{j} \right) 
 & \nonumber\\
 &+ U_{ab} \sum_{j=1}^{N} n_{a,j} n_{b,j} + V_h \sum_{j=1}^{N} (j-L/2)^2 \left(n_{a,j}+n_{b,j} \right) & , 
\label{hub}
\end{eqnarray} 
where $a_j$ and $b_j$ are either bosonic or fermionic field operators acting on a site $j$, with corresponding density operators $n_{a,j}=a_{j}^{\dagger} a_{j}$ and $n_{b,j}=b_{j}^{\dagger} b_{j}$, $t$ is the tunnel constant, $U_{ab}$, $U_{aa}$ and $U_{bb}$ are the on-site interactions. The strength of the on-site interaction is $U_{ab}=U$. If particles of kind $a(b)$ are allowed to interact among them $U_{aa(bb)}=U$, otherwise $U_{aa(bb)}=0$. The strength of the harmonic potential is $V_h$.

The discrete Hamiltonian Eq. \ref{hub} can be solved with the DMRG method \cite{Wall2009,WallNJoP2012,itensor}, and the DMRG results presented in this work are obtained with the open-source codes from L. D. Carr and his group \cite{Wall2009,WallNJoP2012} and with the open-source code from the iTensor project \cite{itensor}, since we are interested in the behavior of the DMRG method rather than a specific code. 
Results from both independently developed codes are consistent, since energies and densities agree in all cases for intermediate values of the interaction parameter, as discussed in Sect. \ref{sec.sisdmrg}.  The interaction parameter that define the intermediate values is found to be the same for both codes. This implies that results discussed in this work are inherent to the DMRG method, and not artifacts of specific codes. Furthermore, results from both codes show that DMRG performs better for ground state than for excited states, which might not be a surprise for specialists in this technique.

\subsection{Variational method}
We will also consider the variational approach proposed in Ref.~\cite{andersenSR2016} which was shown to be highly accurate in estimating the ground state energy of two-component fermion systems up to 6 particles ($N_{\uparrow}=1$ + $N_{\downarrow}=2,...,5$). The method consist of proposing a trial state $\left| \gamma \right\rangle$ which is a superposition of the non-interacting wave function $\left| \gamma_0 \right\rangle$ and the wave function in the strongly interacting limit $\left| \gamma_\infty \right\rangle$, namely 
\begin{equation}
\left| \gamma \right\rangle = \alpha_0 \left| \gamma_0 \right\rangle + \alpha_\infty \left| \gamma_\infty \right\rangle .
\label{varwf}
\end{equation}
Using the Hamiltonian in Eq.~Eq. \ref{hamil}, the variational energy is given by $E = \left\langle \gamma \right| H_c \left| \gamma \right\rangle/\left\langle \gamma \right| \left. \gamma \right\rangle$. Defining $\Delta E = E_\infty - E_0$, the minimization process leads the coefficients and the variational energy to be given by
\begin{eqnarray} 
\frac{\alpha_0}{\alpha_\infty}&= \frac{\Delta E-\left\langle \gamma_0 \right|U\left| \gamma_0 \right\rangle+\sqrt{\left( \Delta E - \left\langle \gamma_0 \right|U\left| \gamma_0 \right\rangle \right)^2+4\left\langle \gamma_0 \right|U\left| \gamma_0 \right\rangle\Delta E \left\langle \gamma_0 \right.\left| \gamma_\infty \right\rangle^2}}{2\left\langle \gamma_0 \right|U\left| \gamma_0 \right\rangle \left\langle \gamma_0 \right.\left| \gamma_\infty \right\rangle} & ,
\label{varcoef} \\
E_{var}&= E_0+ \frac{\Delta E+\left\langle \gamma_0 \right|U\left| \gamma_0 \right\rangle+\sqrt{\left( \Delta E + \left\langle \gamma_0 \right|U\left| \gamma_0 \right\rangle \right)^2-4\left\langle \gamma_0 \right|U\left| \gamma_0 \right\rangle\Delta E \left(1-\lambda^2 \right)}}{2\left(1-\lambda^2 \right)} & ,
\label{vare}
\end{eqnarray}
with $U_{ij} \left| \gamma_\infty \right\rangle=0$. In order to get the right energy behavior at $1/g\to0$, the term $\left\langle \gamma_0 \right.\left| \gamma_\infty \right\rangle^2$ is replaced by $\lambda^2= K \left\langle \gamma_0 \right|U\left| \gamma_0 \right\rangle / (g \Delta E^2)$ in Eq.~Eq. \ref{vare} \cite{andersenSR2016}, with K defined in equation Eq. \ref{slopes}.

The non-interacting wave function is easily calculated once the statistics of each particle is known. In the other limit, the wave function is numerically found with exact diagonalization \cite{DehkharghaniSR2015}, it may be analytically obtained for three or four particles when similar species of particles do not interact with each other \cite{ZinnerEEL2014,Garcia-MarchPRA2014,DehkharghaniPRA2015,LoftTEPJD2015}, or in the most general case the wave function is exactly given up to 30 particles in the strongly interacting limit 
\cite{LoftJoPBAMaOP2016}. Here we show that this variational method works extremely well and 
use it to estimate energies as function of the interaction strength $g$ where numerical or analytical results are not available. Moreover, we compare the wave function Eq. \ref{varwf} with the one obtained using the continuum and discrete descriptions introduced above.

\section{Discrete-to-continuous mapping} \label{sec.dcm}
The beauty in being able to describe the same system from several perspectives is that one can benefit from the power of each method and also avoid their setbacks. However, to exploit this power, we must relate parameters and results from the different approaches. Although there is no trivial way to make this connection, recent efforts have successfully connected parameters from some specific discrete to continuous Hamiltonians using spin models in the strongly interacting case \cite{VolosnievNC2014,DeuretzbacherPRA2014}. Also a recent study has shown how to relate Hubbard models in the continuum to tight-binding lattice models within effective field theory \cite{ValienteFS2015}.

In our case, a glance at the expressions Eq. \ref{hamil} and Eq. \ref{hub} shows that indeed there are many different parameters we would have to connect between the continuous and discrete Hamiltonians. Instead of relating parameters from the Hamiltonians beforehand, as previous works have done, we rather use a way of connecting results from both methods directly. As result, energy spectra and the inter-particle interaction strengths from continuous and discrete descriptions are related through simple mathematical expressions. 

\begin{figure}[!h]%
\subfigure[\ Step 1: Numerical solution of the Hamiltonians given in equationsEq. \ref{hamil} and Eq. \ref{hub}. Inset shows in detail the behavior of $E_d$. \label{ugma}]{\includegraphics[width=0.48\columnwidth]{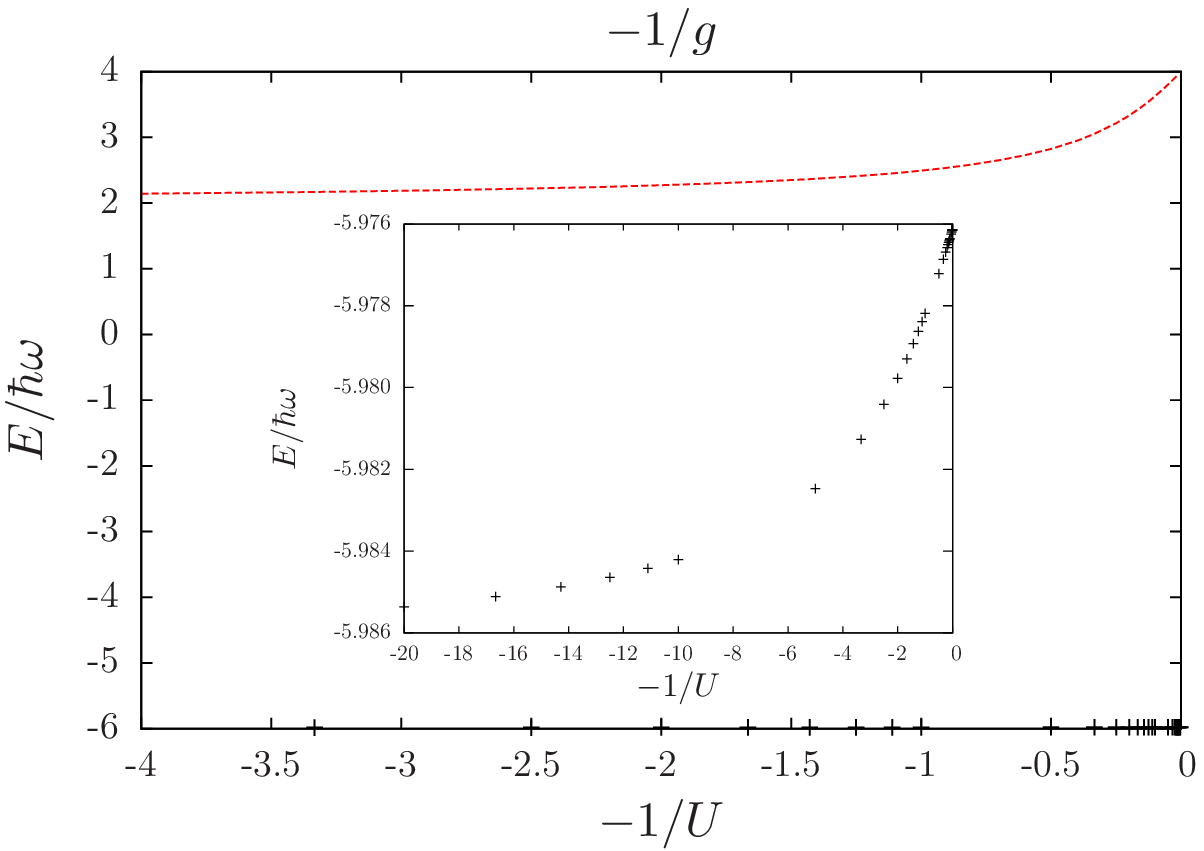}}
\subfigure[\ Step 2: Energies of the discrete Hamiltonian Eq. \ref{hub} calculated with the DMRG method are shifted accordingly to equation Eq. \ref{eshift}. 
\label{ugmb}]{\includegraphics[width=0.48\columnwidth]{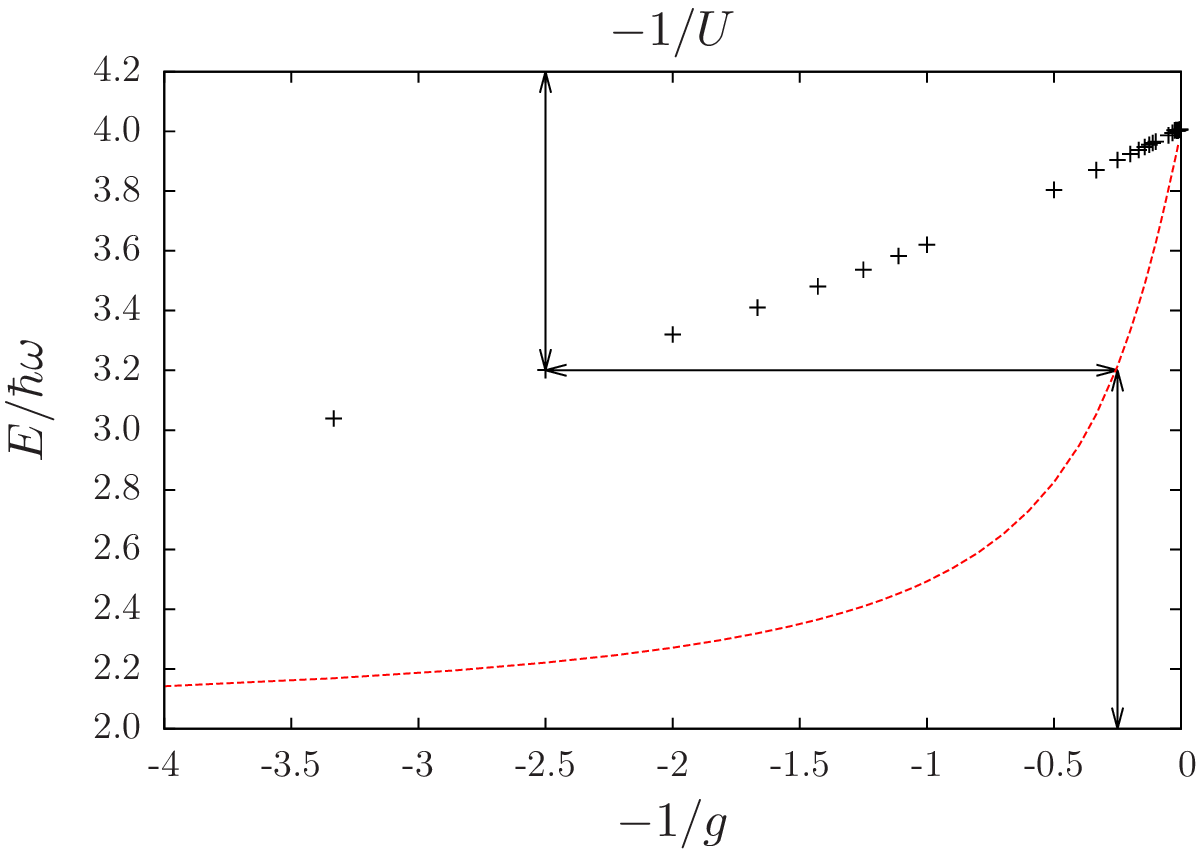}}

\center
\subfigure[\ Step 3: Parameters $U$ and $g$ are connected through $U=0.10291 g$. \label{ugmc}]{
\includegraphics[width=0.48\columnwidth]{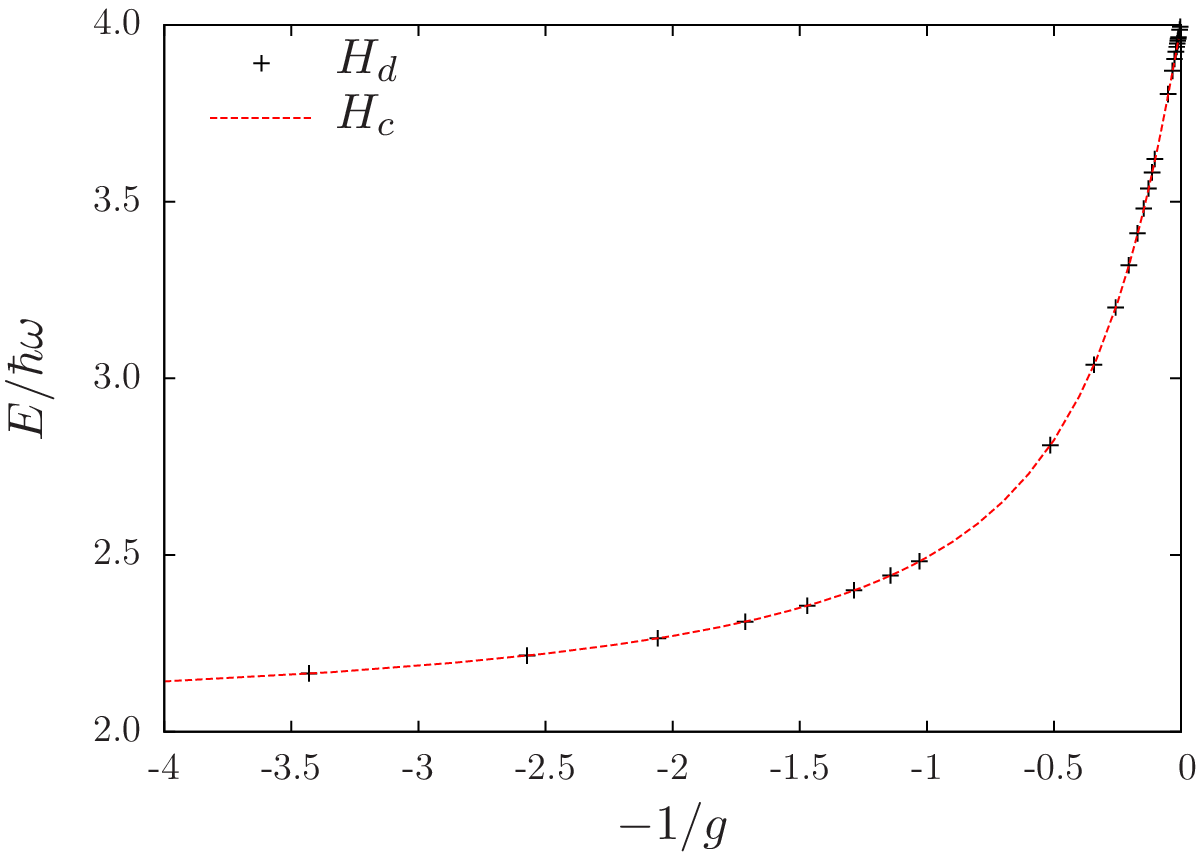}}

\caption{Graphical representation of the steps involved in the discrete-to-continuous mapping. Example for two-component fermion system with $N=3$ ($N_\uparrow=1$ and $N_\downarrow=2$). The parameters of $H_d$ in equation Eq. \ref{hub} are $L=128, V_h/t=7\cdot 10^{-6}, U_{aa}=U_{bb}=0, U_{ab}=U, t=1$. }%
\label{ugm}%
\end{figure}

First of all, the non-interacting part of the Hamiltonians Eq. \ref{hamil} and Eq. \ref{hub} are related. The following procedure establishes a connection between the energy scales of both Hamiltonians, as it connects the ground state one-body energy and the difference between energy levels from both expressions. 
To illustrate the procedure, let us consider a two-component
fermion system with $N=3$ ($N_{\uparrow}=1$ + $N_{\downarrow}=2$). The first step is to find energies as function of the interaction parameter ($g$ for continuous and $U$ for discrete system) as a solutions of the Hamiltonians 
Eq. \ref{hamil} and Eq. \ref{hub}. The results are shown in Fig. \ref{ugma}. Next, the energy calculated from the discrete Hamiltonian Eq. \ref{hub}, labeled $E_d$, has to be shifted by
\begin{equation}
\frac{E}{\hbar \omega}=\frac{E_d-N E_{1p}}{\hbar \omega_d}+N \frac{1}{2} , 
\label{eshift}
\end{equation} 
where $E_{1p}$ is the one particle ground state energy and $\hbar \omega_d$ is the difference between energy levels, both calculated in the discrete model Eq. \ref{hub}, and $N$ is the total number of particles. The shifted result is shown in Fig. \ref{ugmb}. The first term on the right-hand-side of equation Eq. \ref{eshift} comes from the interaction energy, where the numerator removes the ground state energy from the discrete model and the denominator brings the energy levels to the units of the harmonic oscillator energy from equation Eq. \ref{hamil}. The second term adds back the non-interacting ground state energy in the same units as the first term.

The last step is to relate the continuous and discrete interaction strength $g$ and $U$. The horizontal arrow in Fig. \ref{ugmb} points out the explicit relation $E(-1/U)=E(-1/g)\approx 3.2$. From this we find $U=2.5$ and $g=0.2572$ (vertical arrows). The interaction strengths can now be related by $U \delta=g$, with $\delta=0.10291$. We find the same value for the shift parameter 
$\delta$ for any value of the energy. Therefore, by shifting horizontally the discrete curve in Fig. \ref{ugmb} by $\delta$, the results obtained from the continuous and discrete Hamiltonians Eq. \ref{hamil} and Eq. \ref{hub} are the same, as shown 
in Fig. \ref{ugmc}. 
For Fermi-Fermi mixtures, the interaction strengths from continuous Eq. \ref{hamil} and discrete Eq. \ref{hub} descriptions are related by
\begin{equation}
U=0.10291 \ g \ .
\label{dcm}
\end{equation}
Similar relation can be obtained for Bose-Fermi and Bose-Bose mixtures.
The procedure is general and works also well for a higher number of particles even when components have the same population, as we shall see in the next section. Although a solution (generally obtained from numerical calculations) of the continuous Hamiltonian must be known it may not be necessary to fully solve equation Eq. \ref{hamil}, which demands a huge effort. Instead, a variational method such as the one presented in Ref.~\cite{andersenSR2016} could be used as a fast and accurate alternative to full numerical calculations.

\section{Strongly interacting systems with DMRG} \label{sec.sisdmrg}

The density matrix renormalization group (DMRG) method is a very efficient tool to solve the discrete Hamiltonian Eq. \ref{hub}. This technique has been used to solve such Hamiltonians in the low-density limit, $N/L\ll1$, and also in the strongly interacting limit \cite{FangPRA2011}. We will see that in this limit, DMRG calculations fail in finding some observables when the interaction parameter $U$ is taken to be very large in equation Eq. \ref{hub}. 
Investigating how DMRG behave in the strongly interacting limit, we show that it is possible to get observables correctly, given that the parameter $U$ in equation Eq. \ref{hub} is limited to large but not excessive values. This may be benchmarked by
comparing to the analytical results for strongly interacting systems \cite{VolosnievNC2014}. Furthermore, the continuous-to-discrete mapping Eq. \ref{dcm} gives a quantitative meaning to the above sentence ``large but not excessive values".

For the reader interested in reproducing any of the results, here we provide some extra information on the parameters used in the simulations. We have checked that most of the default settings in both Carr's group \cite{WallNJoP2012} and iTensor \cite{itensor} codes did not need to be changed. However, results converged better within $10 - 20$ sweeps and with the bond dimension allowed to grow up to $30$. The discarded weight is in general $<10^{-12}$. Specifically, the minimum cutoff after each SVD operation in the iTensor code was set to $<10^{-13}$.

\subsection{Impurity systems}

\begin{figure}[!htb]%
\centering
\includegraphics[width=0.8\columnwidth]{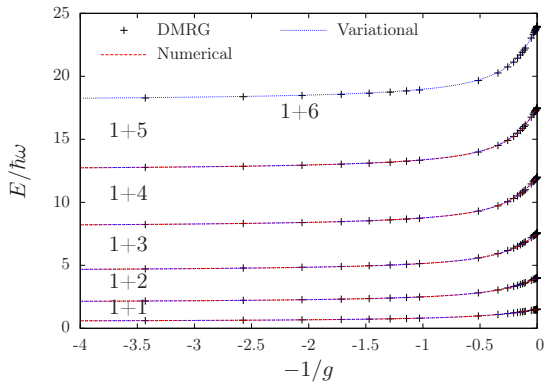}%
\caption{Energy as function of the inverse of the interaction parameter $g$ ($U=0.10291 g$) for a system composed for $N_\uparrow=1$ and $N_\downarrow=1,...6$. ``Numerical" refers to the exact diagonalization of equation Eq. \ref{hamil}, ``DMRG" stands for results from equation Eq. \ref{hub} properly shifted by the discrete-to-continuous mapping given in equation Eq. \ref{eshift} and ``Variational" labels results from equation Eq. \ref{vare}. Results from the three methods agree with accuracy better than $0.3\%$. The parameters of $H_d$ in equation Eq. \ref{hub} are $V_h/t=7\cdot 10^{-6}, U_{aa}=U_{bb}=0, U_{ab}=U, t=1,L=128$ for $N_\downarrow=1,2,3$ and $L=256$ for $N_\downarrow=4,5,6$.}%
\label{Eu1dN}%
\end{figure}

We define impurity systems as those in which a single particle of one internal state interact with a number of particle in a different internal state. Here we consider two-component fermions with $N_\uparrow=1$ and $N_\downarrow=1,\ldots,6$ \cite{GriningNJoP2015,GharashiPRL2013,BugnionPRA2013}. 
The energies of such systems may now be calculated according to the three methods described in 
Sect. \ref{sec.sm} and Fig. \ref{Eu1dN} shows the results. 
First of all, energies from the variational method in equation Eq. \ref{vare} are almost indistinguishable from the numerical exact diagonalization calculation using equation Eq. \ref{hamil} and the accuracy stays within the values given in Ref.~\cite{andersenSR2016}. 
Here we have used the wave function in the strongly interacting limit as described in Refs.~\cite{VolosnievNC2014,VolosnievFS2014,decampNJoP2016}. 

Figure \ref{Eu1dN} further shows that the discrete-to-continuous mapping from Sect. \ref{sec.dcm} is not sensitive to the number of particles. All curves in this figure, representing the ground state energy as function of the interaction parameter for systems with $2\leq N \leq 7$ obey equation Eq. \ref{dcm}. Notice that DMRG calculations are able to reproduce the linear behavior of the energy \cite{VolosnievNC2014,LindgrenNJoP2014} in $-1/g$ as $g\to\infty$. In general, deviations from the mapped DMRG energy in equation Eq. \ref{eshift} to the numerical calculated energies are less than $0.11\%$ for $g<10$ in all cases shown in Fig. \ref{Eu1dN} and no more than $0.3\%$ at the other end, i.e. where $g>10$.

\begin{figure}[!htb]%
\includegraphics[width=1\columnwidth]{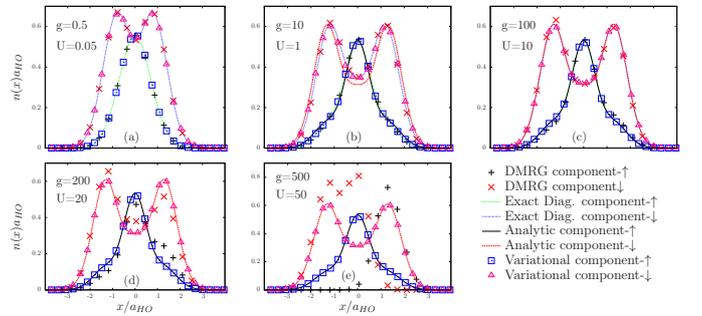}%
\caption{Density profiles of a two-component fermion system with $N_\uparrow=1$ and $N_\downarrow=2$ for a broad range of interaction strengths $g$ ($U=0.10291 g$). Comparison between DMRG, exact diagonalization, variational calculation and analytical result in the strongly interacting limit. Panel (b) shows that the strongly interacting regime has not being numerically achieved at $g=10$ ($U=1$) yet. For $g>100$ ($U>10$) results from exact diagonalization (not shown in panels (d) and (e)) and variational calculation agree with the analytical expression, while DMRG does not perform well. The parameters of $H_d$ in equation Eq. \ref{hub} are $V_h/t=7\cdot 10^{-6}, U_{aa}=U_{bb}=0, U_{ab}=U, t=1,L=128$.}%
\label{densDMRGvsVarVsAna}%
\end{figure}

The scenario changes drastically when we look at other observables such as density profiles. In this case, DMRG calculation nicely agrees with numerical results all the way from the non-interacting limit until the strongly interacting limit is reached in numerical calculations which happens around $g\approx100$, as shown in Fig. \ref{densDMRGvsVarVsAna}(a)-(c). Here the densities from the exact diagonalization of equation Eq. \ref{hamil} are the same as the analytical prediction. 
However, increasing the interaction strength $g$ even more and going towards the exact strongly interacting limit $g\to\infty$, the density profiles obtained from DMRG deviate completely from the known analytical result and have no meaning as in Fig. \ref{densDMRGvsVarVsAna}(d). The extreme case of very large $g$ suggests that DMRG calculations get stuck in a particular state (not necessarily an eigenstate of the system) as shown in Fig. \ref{densDMRGvsVarVsAna}(e), which represents the spatial configurations $\downarrow\downarrow\uparrow$, while the true ground state is composed for a non-trivial combination among the three distinguishable configurations $\downarrow\downarrow\uparrow$, $\downarrow\uparrow\downarrow$ and $\uparrow\downarrow\downarrow$.

The state where the DMRG code is stuck seems to be random, as slightly different parameters can lead to completely different results for $g>100$ $(U>10)$. This can be understood by noticing that the energy spectrum in the strongly interacting limit is highly degenerated and energy levels get closer to each other as $g\to\infty$. Therefore, it is very hard for DMRG methods to identify and isolate the correct state, leading to the results shown in Fig. \ref{densDMRGvsVarVsAna}. 

Since DMRG is being broadly used to solve the discrete model Eq. \ref{hub} in the investigation of continuous systems ($L \to \infty$), we point out that ground state observables, in this case, are only reliable if the interaction parameter stays within $U/t \lesssim 10$. This conclusion does not hold for excited states, as we show in the following.
Analytical or semi-analytical 
\cite{VolosnievNC2014,andersenSR2016} inputs on the strongly interacting limit are then clearly needed and they might also be useful tools to improve DMRG codes in the future.

\subsection{Excited state for $N_\uparrow=N_\downarrow=2$}
\begin{figure}[!htb]%
\center
\includegraphics[width=0.8\columnwidth]{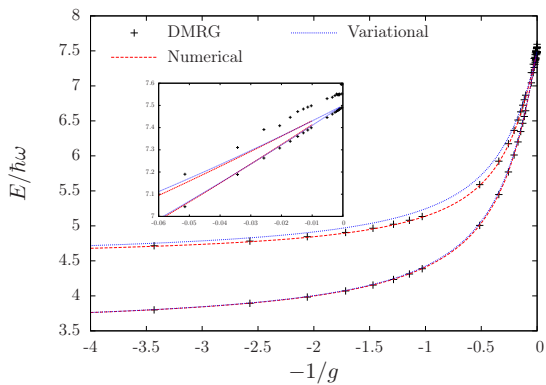}%
\caption{Energy for the ground (bottom) and first excited (top) states as function of the inverse of the interaction parameter $g$ ($U=0.10291 g$) for a system composed for $N_\uparrow=N_\downarrow=2$. ``Numerical" refers to the exact diagonalization of equation Eq. \ref{hamil}, ``DMRG" stands for results from equation Eq. \ref{hub} properly shifted by the discrete-to-continuous mapping given in equation Eq. \ref{eshift} and ``Variational" labels results from Eq. \ref{vare}. The parameters in $H_d$ are $V_h/t=7\cdot 10^{-6}, U_{aa}=U_{bb}=0, U_{ab}=U, t=1,L=128$.}%
\label{Eu2d2}%
\end{figure}

We focus now on two-component fermions with $N_\uparrow=N_\downarrow=2$ \cite{GriningNJoP2015,GharashiPRL2013,GriningPRA2015,CuiPRA2014} and extend the analysis also to the first excited state. DMRG calculation still performs well in finding the ground state energy of the system for any value of the interaction parameter $g$. Results for the first excited state agrees with numerical calculation for small and intermediate values of $g$, but accuracy is lost when the strongly interacting limit is approached as shown in the inset of Fig. \ref{Eu2d2}. This figure further shows that the discrete-to-continuous mapping of equations Eq. \ref{eshift} and Eq. \ref{dcm} does not also depend on the state of the quantum system, given another example for its efficiency and power. 
Finally, we see that the first state energy calculated with the variational method is less accurate, as it would be expected. 

\begin{figure}[!htb]%
\subfigure[\ Ground state for $g=100$ ($U=10$). \label{dens2u+2dA}]{
\includegraphics[width=0.48\columnwidth]{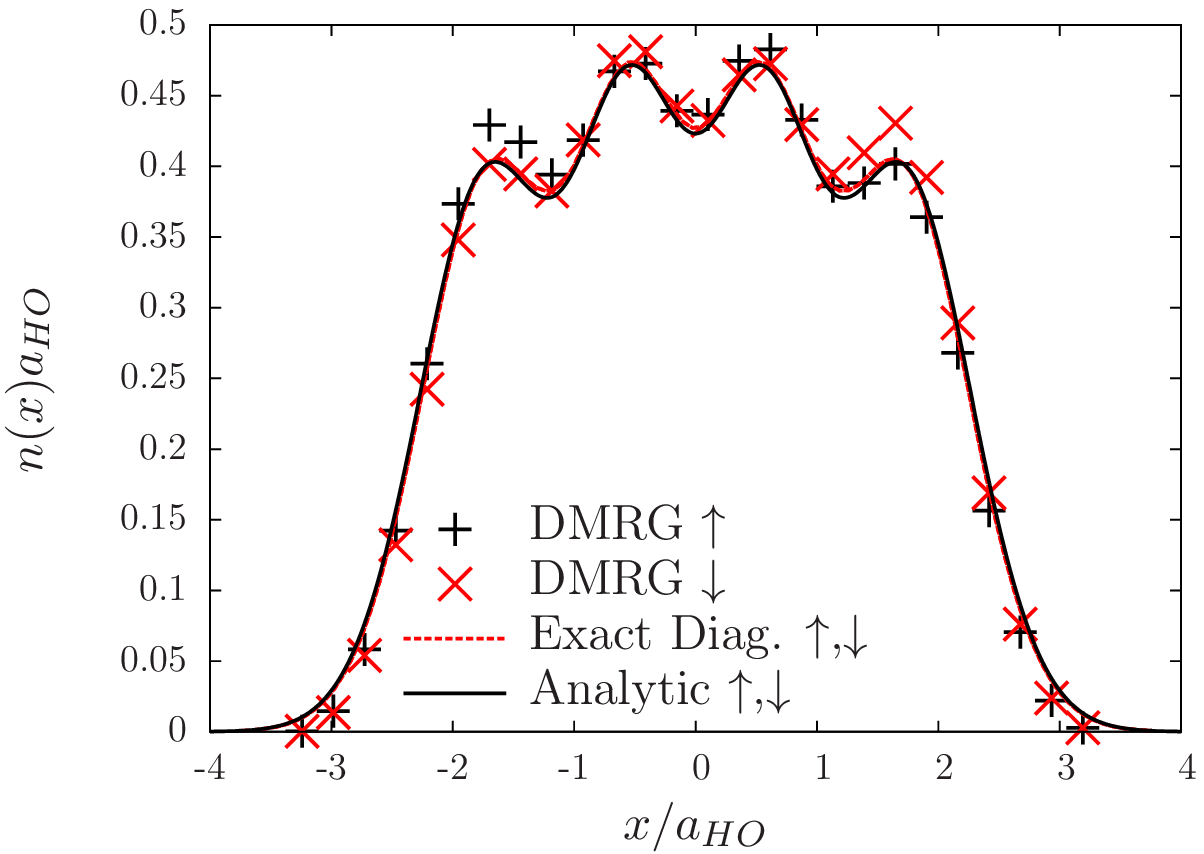}}
\subfigure[\ First excited state for $g=50$ ($U=5$). \label{dens2u+2dB}]{
\includegraphics[width=0.48\columnwidth]{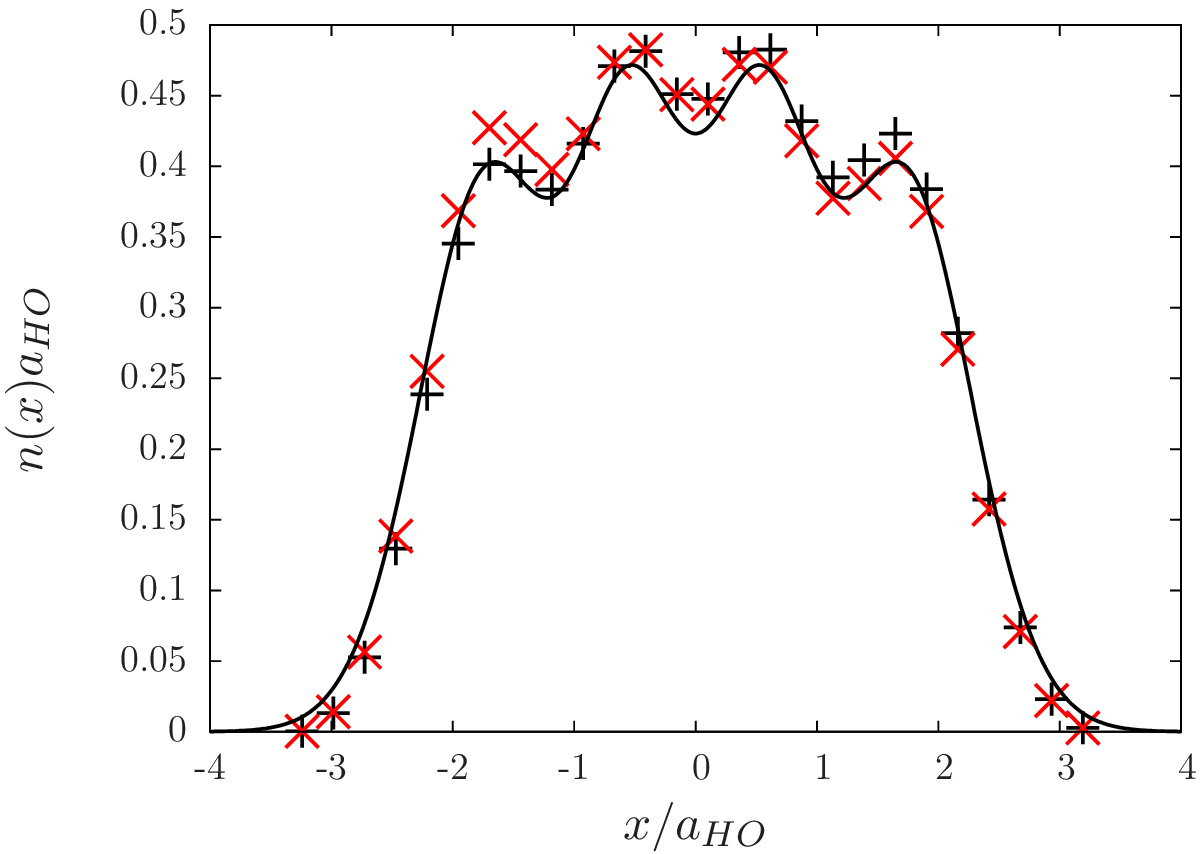}}
\caption{Density profiles for the ground (left) and first excited (right) states of a two-component fermion system with $N_\uparrow=N_\downarrow=2$. Densities for different states are the same in the strongly interacting limit. This limit is numerically achieved at $g\approx100$ for the ground state and at $g\approx50$ for the first excited state. Note that DMRG calculation gives slightly different results for each state. The parameters of $H_d$ in equation Eq. \ref{hub} are $V_h/t=7\cdot 10^{-6}, U_{aa}=U_{bb}=0, U_{ab}=U, t=1,L=128$.}%
\label{dens2u+2d}%
\end{figure}

The density profiles of each component for both states, shown in Fig. \ref{dens2u+2d}, are exactly the same in the strongly interacting limit case where the states are said to be fermionized \cite{LindgrenNJoP2014}. DMRG calculations are able to reproduce the overall behavior of the profiles for both states, however a glance at Fig. \ref{dens2u+2d} also shows that these results have very limited physical meaning, since parity is broken and it is possible to identify the different components in the mixture. Notice that the behavior of the densities calculated from DMRG does not depend on whether the energy of the states is well captured or not. DMRG results for the ground state energy deviate $0.13\%$ from numerical exact calculation for $g=100$ ($U=10$), while the deviation from the first excited energy at $g=50$ (and $U=50$) is three times larger (see Fig. \ref{Eu2d2}). However, Fig. \ref{dens2u+2d} shows that there is basically no difference in accuracy between the profiles for the ground and first excited states, both are inaccurate. 

The overall agreement between densities calculated analytically and with DMRG becomes worse when we look at the pair correlation function Eq. \ref{paircoreq}. This tells us how the different species are spatially organized in the trap and allows us to distinguish the spatial configuration of different states, which is hard to obtain from density profiles alone. For example, looking at Fig. \ref{dens2u+2d} one might argue that DMRG results are as good for the first excited state as they are for the ground state. We now use the pair correlation function to show that this is not true. 

\begin{figure}[!htb]%
\subfigure[\ Ground state. \label{u2d2s0}]{
\includegraphics[width=0.48\columnwidth]{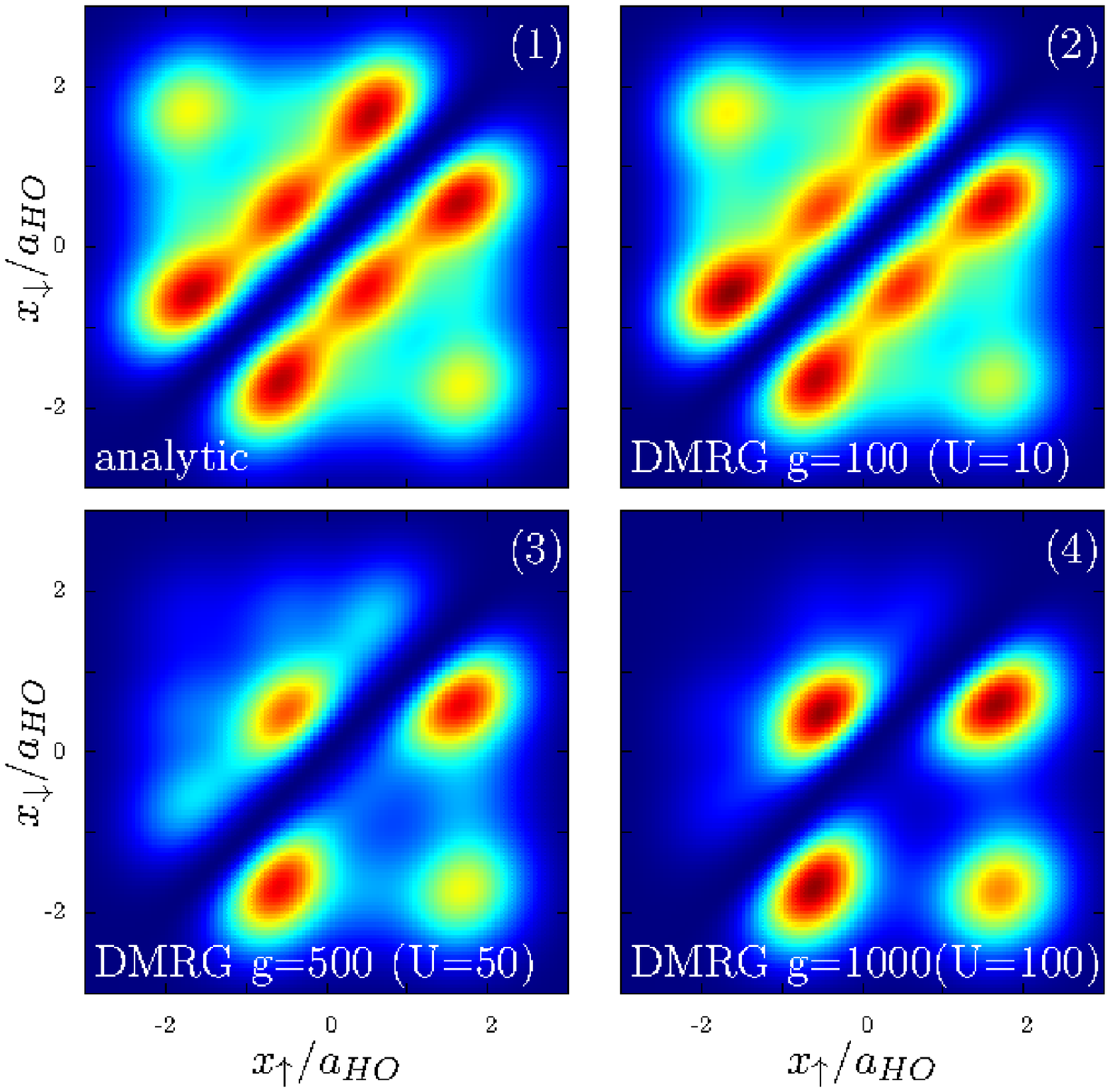}}
\subfigure[\ First excited state. \label{u2d2s1}]{
\includegraphics[width=0.48\columnwidth]{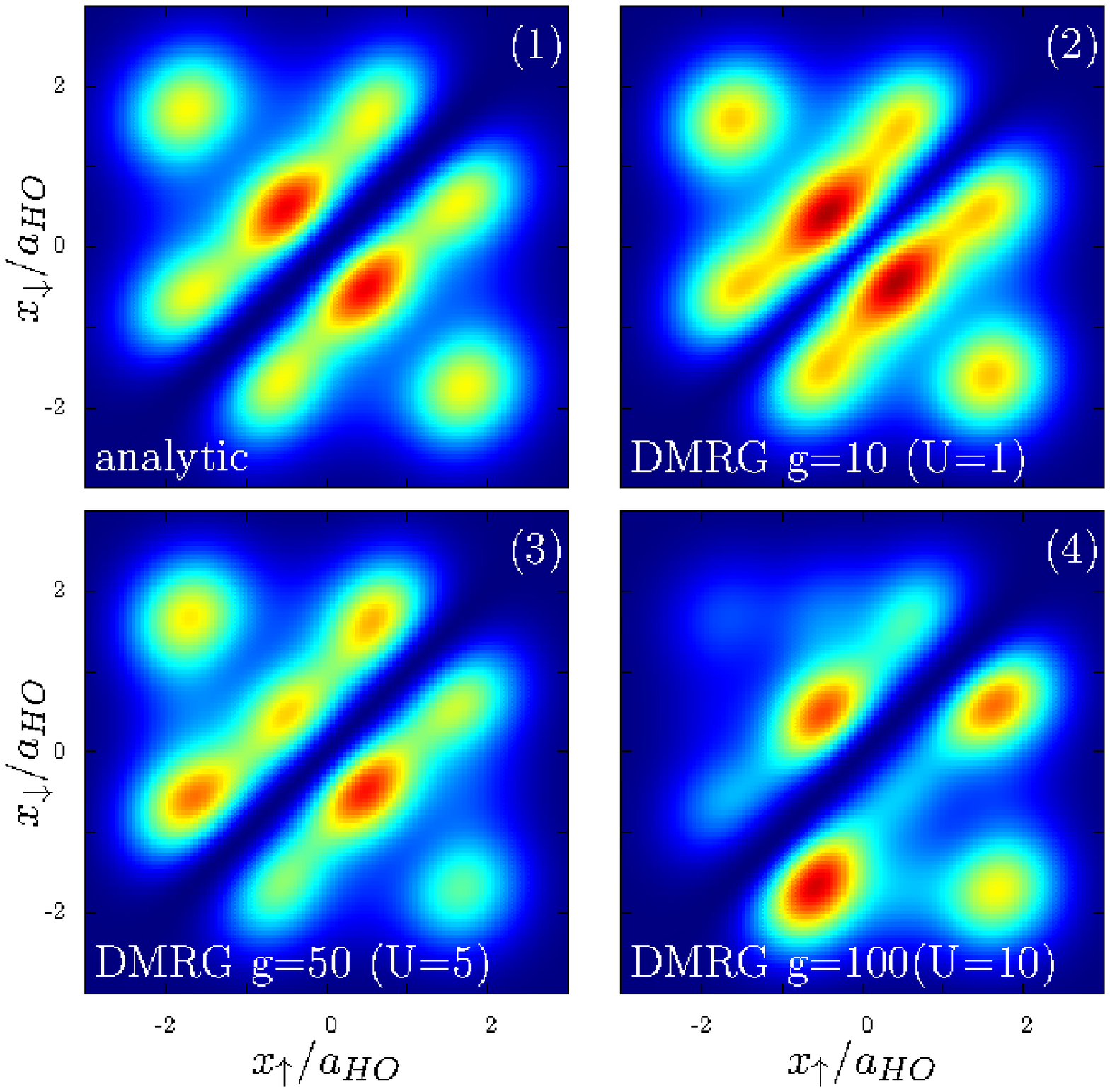}}
\caption{Pair correlation functions of a two-component fermion system with $N_\uparrow=N_\downarrow=2$. Comparison between analytical and numerical results for ground and first excited state in the strongly interacting limit. The parameters of $H_d$ in equation Eq. \ref{hub} are $V_h/t=7\cdot 10^{-6}, U_{aa}=U_{bb}=0, U_{ab}=U, t=1,L=128$.}%
\label{paircorfig}%
\end{figure}

Analytical results for the strongly interacting limit and DMRG results at $g=100 \ (U=10)$ for the ground state pair correlation function are shown in 
Fig. \ref{u2d2s0}. As we have seen for the density profiles in Fig. \ref{dens2u+2d}, DMRG and numerical results agree very well until the strongly interacting limit is reached ($g\approx100$ and $U\approx 10$) from where increasing $g$ either leads to non-physical results or gets the code stuck in a particular state as seen in the bottom panels of Fig. \ref{u2d2s0}. Panel (4) on Fig. \ref{u2d2s0} corresponds to the particles having a spatial configuration of the form
$\downarrow\uparrow\downarrow\uparrow$ which is certainly not the case as the ground state contains a mix of different configurations.

For the first excited state, densities and pair correlation functions resembles the strongly interacting ones at $g\approx10 \ (U\approx 1)$.
However, as seen in the top panels of Fig. \ref{u2d2s1} that display the analytical result in the strongly interacting limit and DMRG result for $g=10$ ($U=1$), the agreement is not as good as for the ground state. Furthermore, the density profile for $g=50$ ($U=5$) 
in Fig. \ref{dens2u+2dB} seems overall similar to the analytical result in Fig. \ref{dens2u+2dA}, but the pair correlation function for the same interaction strength presented in panel (3) of Fig. \ref{u2d2s1} is quite different from the analytical result.

For the first excited state, observables calculated from the discrete Hamiltonian Eq. \ref{hub} in the continuous ($L \to \infty$) and strongly interacting limit ($U \to \infty$) are only reliable if the interaction parameter stays within $U/t \lesssim 1$, which is ten times less than the value found for the ground state observables. Analytical results are able to reach horizons that are hard to be reached numerically and are an essential tool in the understanding of strongly interacting trapped system in 1D.

\section{Conclusion} \label{sec.conc}

We have studied $N$-body system with repulsive short-range interaction in one spatial dimension using three methods to describe the system and find its solutions. The variational method gives accurate results at low computational time and cost and it can be easily implemented once one has the knowledge of the wave function in the strongly interacting limit. Exact diagonalization calculations are standard and efficient techniques employed in the study of continuous systems, which are however limited by the number of particles and demands a great effort for effective implementation. DMRG techniques are arguably the state-of-art method to study discrete systems. An efficient implementation of this advanced technique is demanding, however there are some excellent open source codes available that allow an almost straightforward access. 

While DMRG has been successfully implemented in almost countless studies, we show that it can be challenged by the
presence of very strong interactions. When the on-site interaction parameter is pushed to arbitrary high values, 
DMRG may give meaningless results for both energies and densities. 
We have shown that if one carefully applies DMRG
then it is still possible to obtain reasonable results. 

Although such limitations might be known for some specialists, we have performed a general and detailed study which for the first time quantify ``strong interaction" in this context. Ground state quantities are only reliable until $U/t \approx 10$, value that is ten times less for the first excited state. 
The lesson here is really that DMRG cannot be considered 
a black-box do-all solver for systems with very strong interactions. The large degenerate manifold of states that 
occur in the limit makes it extremely hard for a DMRG code to find the correct solutions for the ground state or 
some specific excited state. The intrinsic variational nature of the DMRG algorithm makes it vulnerable to large
(quasi)-degenerate spaces such as is the case for very strong interactions.

This problem highlights how important analytical knowledge about the strongly interacting limit is. Here 
we have used the method described in \cite{VolosnievNC2014}. This involves a mapping to a 
spin model with local exchange coefficients, which are high-dimensional integrals. 
Fortunately, for external harmonic confinement results up to 30 particles 
have been reported \cite{LoftJoPBAMaOP2016} and this is a sufficiently large particle 
number for most cold atomic gas experiments confined down to a single spatial dimension.
Open source codes are available \cite{LoftCPC2016,DeuretzbacherPRA2016}
from which one may obtain the exact spin model in the case of arbitrary potentials
as well. It would be very interesting to try to combine DMRG with these
analytical results so as to make DMRG much more reliable also in the 
case of very strong interactions.

One way of approaching this is to use the spin models that you get as a starting point directly in a DMRG routine that solves lattice spin models. This would allow one to address many observables and use the fact that DMRG is accurate and can be scaled to larger system sizes for spin models than typical exact diagonalization routines which are exponentially slow for longer spin chains. More generally, one may also consider an approach where one expands the Hamiltonian in a basis set \cite{KollerPRL2016}. In an occupation number basis one may then by appropriate truncation produce lattice models that can be solved using a DMRG routine. It would be very interesting to compare the latter method to the results of the lattice spin model for very strong interactions in order to test how well it performs as one approaches the strongly interacting regime.

\begin{acknowledgement}
This work was supported by the Danish Council for Independent Research DFF
Natural Sciences, the DFF Sapere Aude program, and the Villum Kann Rasmussen
foundation. The authors thank M.~E.~S. Andersen, N.~J.~S. Loft, A.~S. Jensen, 
D.~V. Fedorov, M. Valiente and U. Schollw{\"o}ck for discussions.
\end{acknowledgement}

The analytical results in the strongly interacting limit have been obtained by F. F. Bellotti and N. T. Zinner and A. S. Dehkharghani have performed the exact diagonalization. Variational and DMRG calculation have been performed by F. F. Bellotti and A. S. Dehkharghani. All the authors have contributed in writing and editing the manuscript.

\section*{References}

\providecommand{\newblock}{}

\end{document}